\documentclass[pop,aps,pacs,twocolumn]{revtex4}
\usepackage{graphicx}
\usepackage{natbib}
\usepackage{amsmath}
\usepackage{amssymb}
\usepackage{amsfonts}
\usepackage{amsbsy}
\usepackage{color}
\begin{document}
\title{Mechanism of runaway electron beam formation during plasma disruptions
  in  tokamaks} 
\author{S. S. Abdullaev$^1$, K.H. Finken$^{2}$, K. Wongrach$^2$, M. Tokar$^1$,
 H.R. Koslowski$^1$,  O. Willi$^{2}$, L. Zeng$^3$, and the TEXTOR team}  
\affiliation
{$^1$ Forschungszentrum J\"ulich GmbH,  Institut f\"ur Energie- und
  Klimaforschung - Plasmaphysik, D--52425 J\"ulich, Germany \\
  $^2$ Institut f\"ur Laser- und Plasmaphysik, Heinrich-Heine 
  Universit{\"a}t D{\"u}sseldorf, Germany \\ $^3$ Institute of Plasma
  Physics, Chinese Academy of Sciences, 230031 Hefei, China} 

\begin{abstract}
A new physical mechanism of formation of runaway electron (RE) beams
during plasma disruptions in tokamaks is proposed. The plasma disruption is
caused by a strong stochastic magnetic field formed due to nonlinearly excited
low-mode number magnetohydrodynamic (MHD) modes. It is conjectured that the
runaway electron beam is formed in the central plasma region confined inside
the intact magnetic surface located between $q=1$ and the closest low--order
rational magnetic surfaces [$q=5/4$ or $q=4/3$, \dots]. It results in
that runaway electron beam current has a helical nature with a predominant
$m/n=1/1$ component. The thermal quench and current quench times are estimated
using the collisional models for electron diffusion and ambipolar particle
transport in a stochastic magnetic field, respectively. Possible mechanisms for
the decay of the runaway electron current owing to an outward drift electron
orbits and resonance interaction of high--energy electrons with the $m/n=1/1$
MHD mode are discussed.   
\end{abstract}
\maketitle

The runaway electrons (REs) generated during the disruptions of tokamak
plasmas may reach a several tens of MeV and may contribute to the significant
part of post--disruption plasma current. The prevention of such RE beams is of
a paramount importance in future tokamaks, especially in the ITER operation,
since it may severely damage a device wall
\cite{Wesson_etal_89,Gill_93,Schuller_95,Gill_etal_00,Papp_etal_13}. \par  

The mitigation of REs by massive gas injections (MGI) and externally
applied resonant magnetic perturbations (RMPs) have been extensively 
discussed in literature (see, e.g.,
Refs. \cite{ITER:2007_3,Hollmann_etal_10,Lehnen_etal_11} and 
references therein). However, no regular strategy to solve this problem has
been developed because up to now the physical mechanisms of the formation of
REs during plasma disruptions are not well understood. In spite of the
numerous dedicated experiments to study the problem of runaway current
generation during plasma disruptions in different tokamaks (see, e.g., 
\cite{Forster_etal_12b,Zeng_etal_13,Wongrach_etal_14,Chen_etal_13,%
Plyusnin_etal_06,Lehnen_etal_11,Hollmann_etal_10,Commaux_etal_11,%
Hollmann_etal_13}) no clear dependence of RE formation on plasma
parameters has been established. These numerous experiments show the complex
nature of plasma disruption process especially the formation of RE
beams. \par   

One of the important features of the formation of RE beams is the
irregularity and variability of the beam parameters from one discharge to
another one. This is an indication of the sensitivity of RE beam formations on
initial conditions which is the characteristic feature of nonlinear processes, 
particularly, the chaotic system. Therefore one expects that {\em ab initio}
numerical simulations of the RE formation process may not be quite productive
to explain it because of complexity of computer simulations of nonlinear
processes \cite{Kadanoff_04}. The problems of numerical simulations of plasma
disruptions is comprehensively discussed in \cite{Boozer_12}.   \par      

In this work we propose a new physical mechanism of formation of RE beams
during plasma disruptions in tokamaks. It is based on the analysis of numerous
experimental results, mainly obtained in the TEXTOR tokamak and the ideas of
magnetic field stochasticity \cite{Abdullaev:2014}. The mechanism explains
many features of plasma disruptions accompanied by RE generations. \par  
 
\textbf{Main conjectures.}
It is believed that the plasma disruption starts with the excitation of
magnetohydrodynamic (MHD) modes with low poloidal $m$ and toroidal $n$
numbers, ($m/n= 1/1, 2/1, 3/2$, $5/2, \dots$) that lead to a large--scale
magnetic stochasticity (see, e.g.,
\cite{Kadomtsev_84,Wesson:2004,Kruger_etal_05,Izzo_etal_12} and 
references therein). The heat and particle transports in the strongly chaotic
magnetic field causes the fast temperature drop and ceases the plasma
current. This process depends on the structure of the stochastic magnetic
field which depends on the spectra of magnetic perturbations and on the safety
factor profile $q(\rho)$, ($\rho$ is the minor radius of the magnetic
surface). At certain conditions the stochastic magnetic field may not extend
up to the central plasma region due to the creation of the outermost intact
magnetic surface $\rho_c$. The electrons confined by this magnetic surface are
accelerated by the toroidal electric field induced by the current decay from 
the outer plasma region, which leads to the formation of the RE beam. 
The initial RE current $I_p^{(RE)}$ is mainly determined by the pre-disruption
plasma current distribution $I_p(\rho)$ confined by the outermost intact
magnetic surface $\rho_c$, i.e., $I_p^{(RE)} \approx I_p(\rho_c)$. \par

The lifetime of the RE beam mainly depends on two effects: the outward
drift of RE orbits induced by the toroidal electric field $E_\varphi$
\cite{Guan_etal_10,Abdullaev_15} and 
the resonant interactions of REs with helical magnetic perturbations.
The first one is responsible for the smooth decay of the RE current, while the
second one cause the sudden RE losses. The outward
drift velocity $v_{dr}$ is determined by $E_\varphi$ and the RE current,
$v_{dr}\propto E_\varphi/I_p^{(RE)} \propto E_\varphi/\rho_c^2$
\cite{Guan_etal_10,Abdullaev_15}. The most stable of the RE beams are expected
to form when the corresponding drift velocity is lowest and the low--order
rational surfaces within the RE beam are absent or one. \par   
 
Consider, for example, the pre-disruption plasma with a monotonic safety factor
profile $q(\rho)$ with $q(0)<1$. Then the \emph{most stable RE beam can be
formed when the outermost intact magnetic surface is located between magnetic
surface $q=1$ and the nearest low--order rational surfaces} $q=5/4$ [or
$q=4/3$, \dots]. It occurs at the sufficiently small amplitude of the $m/n=1/1$
mode. There is only one rational magnetic surface $q=1$ within the RE beam
that is resonant to the large--scale magnetic perturbations, particularly, to
the RMPs. Such RE beams are relatively stable since low--energetic REs (up to
10--15 MeV) are not destabilized due to absence of a large scale
stochasticity. The loss of REs mainly occurs due to the outward drift of RE
orbits and the stochastic instability of high--energetic REs due to the
interactions of high--mode harmonics of the $m/n=1/1$ mode of magnetic
perturbations.  \par  

In the case of plasma disruptions with $q(0)>1$ the intact magnetic
surface $\rho_c$ would be smaller while the toroidal electric field
$E_\varphi$ would be larger than in the ones with $q(0)<1$. Due to the large
outward drift velocity $v_{dr}$ such RE beams would cease faster. \par      

\begin{figure} 
\centering  
\includegraphics[width=0.5\textwidth]{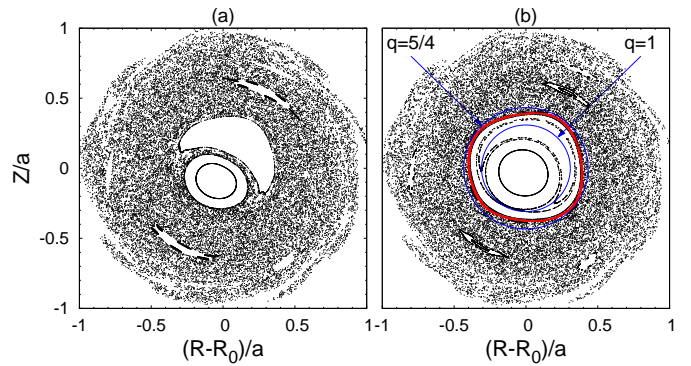}  
\caption{(color online) Poincar\'e sections of magnetic field lines in a
  pre--disruption plasma caused by several $m/n$ MHD modes, ($n=1,2,3;
  m=1,\dots 8$): (a) all mode amplitudes $B_{mn}$ are equal; (b) the amplitude
  $B_{11}$ of the $m/n=1/1$ mode is four times smaller than $B_{mn}$. The
  safety factor at the magnetic axis $q(0)=0.8$ and at the plasma edge
  $q(a)=4.7$.}    
\label{disrup_plasma_sect1}
\end{figure}

The two possible distinct generic structures of a stochastic magnetic field
before the current quench with the RE-free discharge and with the RE discharge
are shown in Figs.~\ref{disrup_plasma_sect1} (a) and (b) by the Poincar\'e
sections of magnetic field lines. It is assumed that the perturbation magnetic
field contains several low--mode number $m/n$ MHD modes with equal amplitudes
$B_{mn}$: (a) the amplitude $B_{11}$ of the $m/n=1/1$ mode is equal to others;
(b) $B_{11}$ is four times smaller than the amplitudes of other modes. As seen
from Fig.~\ref{disrup_plasma_sect1} (a) for the large amplitude of the
($m/n=1/1$) mode the stochastic magnetic field extends up to the central
plasma region destroying the separatrix of the $m=n=1$ island. For the
low--amplitude of the ($m/n=1/1$) mode shown in Fig.~\ref{disrup_plasma_sect1}
(b) the stochastic magnetic field does not reach the $q=1$ magnetic
surface. The last intact drift surface (red dots) is located between the
resonant surfaces $q=1$ and $q=5/4$ (blue curves).  \par     
The existence of an intact magnetic surface and its location depends on the 
radial profile of the safety factor and on the spectrum of magnetic
perturbations. The latter sensitively depend on the plasma disruption
conditions and vary unpredictably from one discharge to another during plasma
disruptions. This makes RE formation process unpredictable and may
explain a shot--to--shot variability of the parameters of RE beams.  \par   

\textbf{Experimental evidences}.
This conjecture on the mechanism of RE beam formation agrees with the
important features of the experimental observations in the TEXTOR tokamak. In
the experiments the plasma disruptions were triggered by gas injections (see,
e.g.,
\cite{Forster_etal_12b,Zeng_etal_13,Wongrach_etal_14,Bozhenkov_etal_08}): the 
disruptions with REs were triggered by argon (Ar) injection and the
RE--free disruptions with Ne injection. The injection of these gases
may finally give rise to different spectra of amplitudes of MHD  modes. One
can expect that the amplitude of the $m/n=1/1$ MHD mode excited by the He/Ne
injection is higher than in the case of Ar gas injection. \par 

\begin{figure} 
\includegraphics[width=0.45\textwidth]{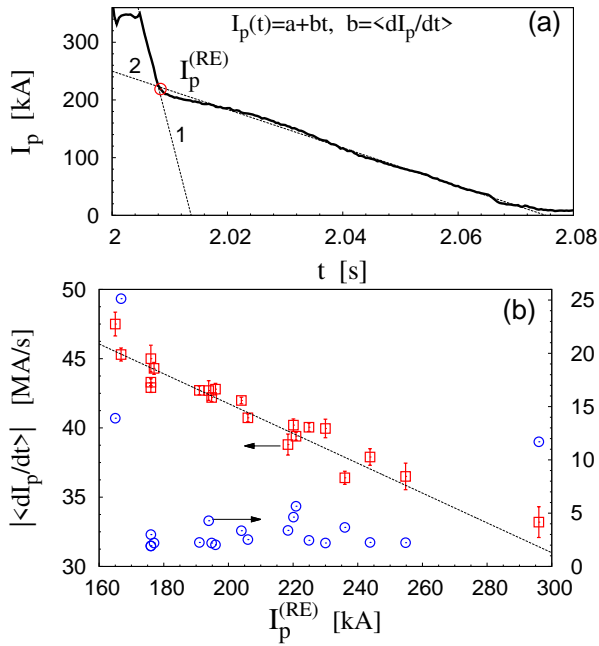} 
\caption{(color online) (a) Typical time evolution of the plasma current with
  RE current. The average current decay rates $\langle dI_p/dt \rangle$ at the
  CQ and the RE plateau stages are determined by fitting with a linear
  function $I_p(t)=a+bt$. Symbol \textcolor{red}{$\odot$} corresponds to the
  plasma current $I_p^{(RE)}$ at the initial stage of the RE 
  plateau. (b) The decay rates $|\langle dI_p/dt \rangle|$ versus
  $I_p^{(RE)}$. Symbols \textcolor{red}{$\boxdot$} correspond to the CQ rate
  (l.h.s. axis), and \textcolor{blue}{$\odot$} $-$ the RE plateau
  (r.h.s. axis). }     
\label{Ip_REdecay}
\end{figure}

\begin{figure} [t]
\centering  
\includegraphics[width=0.49\textwidth]{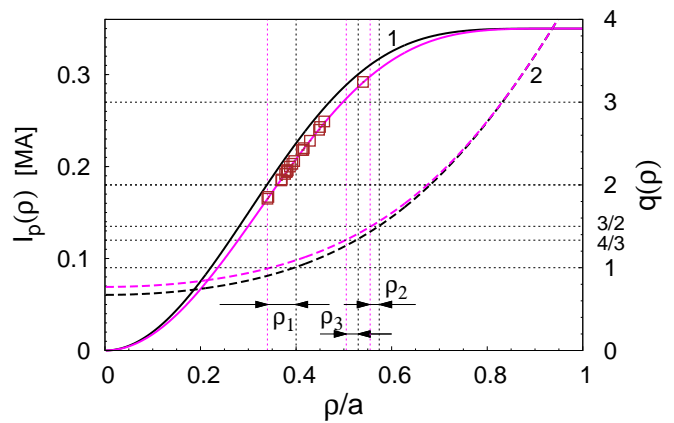} 
\caption{(color online) Radial profile of the plasma current $I_p(\rho)$
  (solid curves 1 on l.h.s. axis) and the corresponding safety factor profile
  $q(\rho)$ (dashed curves 2 on r.h.s. axis). The rectangular (red) dots
  correspond to the experimentally measured values of $I_p^{(RE)}$ for several
  TEXTOR discharges. The plasma parameters are $I_p=350$ kA, $B_0=2.4$ T,
  $R_0=1.75$ m, and $a=0.46$ m. The values of $q(0)$ are 0.7 (solid black
  curves) and 0.8 (dashed magenta curves), respectively. The radii $\rho_1$,
  $\rho_2$, and $\rho_3$ are the positions of the rational magnetic surfaces
  $q(\rho_1)$=1, $q(\rho_2)$=3/2, and $q(\rho_3)$=4/3, respectively. }     
\label{disrup_plasma}
\end{figure}

The plasma current decay in the current quench (CQ) and the RE plateau
regimes for all discharges is well approximated by the linear function of time
$I_p=I_{p0}+bt$, with the average CQ rate $b=\langle dI_p/dt
\rangle$ as shown in Fig.~\ref{Ip_REdecay} (a). The current decay rates
$|\langle dI_p/dt \rangle|$ in the CQ stage and the RE plateau stage versus
the initial RE current $I_p^{(RE)}$ for a number discharges are plotted
Fig.~\ref{Ip_REdecay} (b).  The plausible radial profiles of $I_p(\rho)$ and
the corresponding safety factor $q(\rho)$ are plotted for the two values of
$q(0)$ in Fig.~\ref{disrup_plasma}. \par 

\emph{Existence of the finite interval of the initial RE currents $I_p^{(RE)}$}.
Since $\rho_c$ is located between the magnetic surfaces $\rho_1$
and $\rho_3$ corresponding to $q(\rho_1)=1$ and $q(\rho_3)=4/3$, the RE
current $I_p^{(RE)}$ should take values in the finite interval. This
expectation is supported by the experimental values of the plasma current
$I_p^{(RE)}$ as seen from Fig.~\ref{Ip_REdecay} (a) and (b). These values of
$I_p^{(RE)}$ also lie in the region between the resonance magnetic surfaces
$q(\rho_1)=1$ and $q(\rho_3)=4/3$ [or $q(\rho_2)=3/2$] as 
shown in Fig.~\ref{disrup_plasma} where the radial profile of the
pre--disruption equilibrium plasma current $I_p(\rho)$ (curve 1) and the
corresponding safety factor profile $q(\rho)$ (curve 2) are plotted. \par 

The average values of $\langle|dI_p/dt| \rangle$ for almost all discharges
are confined in the interval (2.2, 5.6) MA/s, i.e., in one order lower than
the current decay rate in the CQ stage. The values of $I_p^{(RE)}$ are
in the range between 170 kA and 260 kA (see Fig.~\ref{Ip_REdecay} (b)). These
values of $\langle|dI_p/dt| \rangle$ and $I_p^{(RE)}$ are close to the ones
observed in the similar experiments in the DIII-D tokamak (see, e.g.,
\cite{Hollmann_etal_13}). \par  

As seen from Fig.~\ref{Ip_REdecay} (a) there are untypical discharges with 
the highest and lowest values of $I_p^{(RE)}$ that correspond to $\rho_c$ at
the borders of region $\rho_1< \rho < \rho_3$. For these discharges the
CQ rates $\langle|dI_p/dt|\rangle$ take highest or lowest values. The RE
current decay rates of these discharges take the highest values. They have the
shortest duration time of RE currents. One expects that the presence of
several low--order $m/n=4/3$, $m/n=3/2$, and $m/n=1/1$ 
resonant magnetic surfaces within the RE beam for the discharge with the
highest $I_p^{(RE)}$ may lead to excitations of the corresponding MHD
modes. The interactions of these modes may lead to the quick loss of REs due
to the formation of a stochastic zone at the edge of the RE beam. \par   

\emph{Dependence on the level of magnetic perturbations}.
The existence of the intact magnetic surface $\rho_c$ between the $q=1$ and
$q=4/3$ rational magnetic surfaces and its location depends on the level
magnetic perturbation $\epsilon_{MHD}$ (more exactly on the spectrum
$B_{mn}$). With increase of $\epsilon_{MHD}$ the radius $\rho_c$ shrinks and
it can be broken at the certain critical perturbation level 
$\epsilon_{cr}$. It leads to the total destruction of confinement of electrons
and ions. This is in agreement with experimental observations of the existence
of critical magnetic perturbations from which on runaway beams are not
generated \cite{Zeng_etal_13}. \par 

The shrinkage of $\rho_c$ with increasing the magnetic perturbation
$\epsilon_{MHD}$ leads to the decrease of the RE current $I_p^{(RE)}$ since
$I_p^{(RE)} \approx I_p(\rho_c)$. On the other hand if one assumes that the plasma
current decay is caused by the radial transport of particles in the stochastic
magnetic field, the CQ rate $dI_p/dt$ should be proportional to the square
of the magnetic perturbation level $\epsilon_{MHD}$, $|\langle dI_p/dt \rangle|
\propto \left|\epsilon_{MHD}\right|^2$. Therefore, one expects that to the
higher values of $|\langle dI_p/dt \rangle|$ correspond the lower values of
the RE current $I_p^{(RE)}$. This expectation is in agreement with the
experimental measured values of these quantities presented in
Fig.~\ref{Ip_REdecay} (b). 

The formation of the RE beam inside the intact magnetic surface can be also
confirmed by the spatial profiles of the synchrotron radiation of high--energy
REs with energies exceeding 25 MeV. One observes that the radiation is
localized within a finite radial extent in the central plasma region. \par    

\textbf{Estimations of thermal quench (TQ) and CQ times.} 
The strong radial transport along the stochastic magnetic field lines causes
the losses of heat and plasma particles from the stochastic zone. The TQ
can be explained by the fact that the anomalously large heat
transport in a stochastic magnetic field is mainly determined by the electron
diffusion. The CQ is determined by the particle transport in
a stochastic magnetic field and  has an ambipolar nature. Using the
collisional test particle transport model in a stochastic magnetic field
\cite{Abdullaev_13} we estimated the heat conductivity $\chi_r(\rho)$ and   
the ambipolar diffusion coefficient $D_p$ of particles. \par   

For typical magnetic perturbations and pre-disruption plasma temperatures
($0.5 \div 1.0$~keV) the magnitude of $\chi_r(\rho)$ has the order of several
$10^2$ m$^2$/s. The characteristic heat diffusion time $\tau_H= a^2/2\chi_r$
is of the order of $10^{-4}$ s that agrees with the experimentally observed
times. The quantitative analysis based on the numerical solution of the heat
diffusion equation also gives similar values for $\tau_H$. \par   

The ambipolar particle transport in a stochastic magnetic field is strongly
collisional due to the low plasma temperature (from 5 eV to 50 eV) after the
TQ. At these plasma temperatures the corresponding diffusion time $\tau_p =a^2
/ D_p$ of particles changes from 1 s to 0.3 s. Since the diffusion coefficient
$D_p \propto B_{mn}^2$ and therefore $\tau_p \propto B_{mn}^{-2}$, then
$\tau_p$ can be reduced to one order smaller value for a three times larger
perturbation than in Fig.~\ref{disrup_plasma_sect1}. This timescale is still
much longer than the experimental values. However, the collisional model does
not takes into account the effect of the toroidal electric field. One
expects that the acceleration of electrons and ions by the toroidal electric
field increases the radial transport of particles. To include this 
effect in the collisional model one can assume that the effective temperature
of the plasma is higher than the measured one. The particle diffusion time
$\tau_p$ at the effective temperature 2 keV is about $8\times 10^{-3}$ s. 
This timescale gives the average current decay rate $dI_p/dt \approx
I_p/\tau_p= 0.35/(8.0 \times 10^{-3}) \approx 44.0$ MA/s which is order of
the experimental measured one given in Fig.\ref{Ip_REdecay} (b).   \par   

In general the transport of heat and particles in the presence of RMPs 
is a three--dimensional problem. Particularly, a stochastic magnetic field
with the topological structures like the ones in
Figs.~\ref{disrup_plasma_sect1} leads to poloidally and toroidally localized
heat and particle deposition patterns on the  wall (see, e.g.,
\cite{Kruger_etal_05}) similar to those in ergodic divertor tokamaks (see,
e.g., \cite{Abdullaev:2014}).  \par  

\textbf{RE beam evolution}.
From the described scenario of plasma disruption it follows that a typical
runaway beam current is localized inside the area enclosed by the last intact
magnetic surface. In general the distribution of the current density $j$ would
depend not only on the radial coordinate $\rho$ but also vary along the
poloidal $\theta$ and the toroidal $\varphi$ angles due to the presence of the
($m/n=1/1$) magnetic island. This agrees with the analysis of numerous
disruptions in the JET tokamak \cite{Gerasimov_etal_14}. One can assume that
the radial profiles of the RE current density averaged along poloidal and
toroidal angles are almost uniform. This gives the value of the safety factor
at the beam axis $q(0)$ is less than unity. This assumption is supported by a
number of experimental measurements of the current profile after the sawtooth
crashes in the TEXTOR, the TFTR, and JET tokamaks
\cite{Soltwisch_etal_87a,Yamada_etal_94,SoltwischKoslowski_95,%
ORourke_91,Koslowski_etal_96,SoltwischKoslowski_97}.    \par 

\begin{figure} [t]
\centering  
\includegraphics[width=0.5\textwidth]{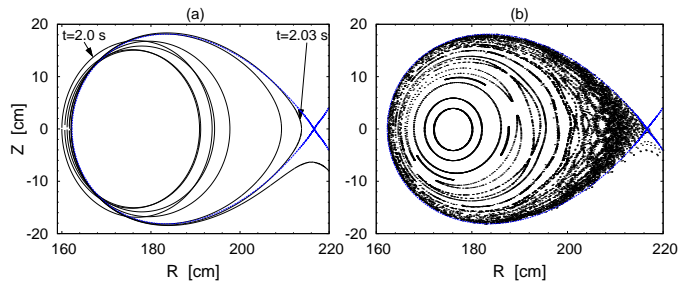}
\caption{(a) Evolution of a RE orbit in the ($R,Z$)-plane for the TEXTOR
  discharge \# 117527. (b) Poincar\'e section of RE orbit of energy 11.7 MeV in
  the ($R,Z$)-plane. The plasma current $I_p=50$ kA.  } 
\label{Ip_orbits}
\end{figure}

The toroidal electric field accelerates electrons to higher energies. With
increasing electrons energy their orbits drifts outwardly 
\cite{Guan_etal_10,Abdullaev_15} and eventually hit the wall. It is
illustrated in Fig.~\ref{Ip_orbits} (a). This effect may be one of mechanisms
of slow RE current decay. Calculations show that the outward drift velocity
$v_{dr}$ is of the order of a few m/s for typical discharges in TEXTOR. The RE
current decay rate $dI_p/dt$ due to outward drift RE orbits can be roughly
estimated as follows. This loss mechanism is mainly caused by the shrinkage of
the beam radius $a$. The rate of such a shrinkage $da/dt$ is of the order of
the average outward velocity $v_{dr}$. Since $I_p\propto a^2$, we have $dI_p/dt
\propto (2I_p/a) da/dt=  (2I_p/a) v_{dr}$. For the typical values of
$I_p\approx 0.2 MA$, $a \approx 0.2$ m, and $v_{dr}\sim 1$ m/s one has
$dI_p/dt \approx 4$ MA/s. This estimation is in the order of the experimentally
measured average decay rate of the runaway current plotted in
Fig.~\ref{Ip_REdecay} (b). \par  

The effect of magnetic perturbation on RE beams depends on their safety factor
profile $q$. The latter varies in the interval $[q(0)<1, q(a)]$ 
with its edge value $q(a)$ less than $3/2$ [or 4/3, 5/3]. Such a RE beam is 
relatively stable to the effect of magnetic perturbations. The single
$m/n=1/1$ mode does not create the stochastic layer at the beam edge for REs
with energies up to several MeVs since their drift surfaces are close to
magnetic surfaces. With increasing the energy of electrons the drift surfaces
strongly deviates from magnetic ones and thus creates the perturbation
harmonics with higher mode numbers $m>1$. The interactions of several
resonance modes of perturbations may form the stochastic zone at the beam edge
which leads to fast RE losses as illustrated in Fig.~\ref{Ip_orbits} (b). This
process may explain the sudden RE current drop accompanied by magnetic
activity and RE bursts observed in experiments (see, e.g.,
\cite{Gill_etal_00,Forster_etal_12b}). \par   

{\em Summary}. Based on the analysis of numerous experimental data obtained in
the TEXTOR tokamak we have proposed the mechanism of RE beam formation
during the plasma disruption. The plasma disruption starts due to a
large--scale magnetic stochasticity caused by nonlinearly excited of MHD modes
with low $(m,n)$ numbers ($m/n= 1/1, 2/1, 3/2$, $5/2, \dots$). The RE beam is
formed in the central plasma region confined by the intact magnetic
surface. Its location depends on the safety factor profile $q(\rho)$ and the
spectrum of MHD modes. In the cases of plasmas with the monotonic profile of
$q(\rho)$ and at sufficiently small amplitude of the $m/n= 1/1$ mode  the
most stable RE beams are formed by the intact magnetic surface located between
the magnetic surface $q=1$ and the closest low--order rational 
surface $q=m/n>1$ ($q=5/4$, $q=4/3$, or $q=3/2$).  \par 

This mechanism reproduces well the essential features of the measurements.
Particularly, the TQ and the CQ are determined by the strong
electron diffusion and ambipolar transport of particles in a stochastic
magnetic field, respectively. The slow decay of the RE current is due to the
outward drift of RE orbits induced by a toroidal electric field, and the
spiky quick decay of REs is due to resonant interaction of high--energy REs
with the $m/n=1/1$ MHD mode. The effect of external resonant magnetic
perturbations on low-energy electrons (up to 5-10 MeV) is weak and does not
cause their loss. This is in agreement with the recent experiments in the
TEXTOR tokamak \cite{Koslowski_etal_14}. 
The detailed description of the mechanism of RE formation and the evolution of
RE current based on the analyses of experimental observations  will be given
in a separate publication \cite{Abdullaev_etal_15}. 

\emph{Acknowledgments}. The authors gratefully acknowledge valuable
discussions with W. Biel, S. Brezinsek, O. Marchuk, Ph. Mertens, D. Reiser,
D. Reiter,  A. Rogister, and U. Samm. Authors also thank Ph. Mertens for
improving the English. 


\end{document}